\documentclass[[12pt,thmsa]{article}
\usepackage{amssymb}

\makeatletter \@addtoreset{equation}{section} \makeatother

\def\ftoday{{\sl {Le \number\day \space\ifcase\month
\or janvier\or f\'evrier\or mars\or avril\or mai \or juin\or
juillet\or ao\^ut\or septembre\or octobre \or novembre \or
d\'ecembre\fi\space \number\year}}}
\def\ptoday{{\sl {\number\day \space de\space \ifcase\month
\or janeiro\or fevereiro\or mar{\c c}o\or abril\or maio \or
junho\or julho\or agosto\or setembro\or outubro \or novembro \or
dezembro\fi\space de\space \number\year}}}
\def\gtoday{{\sl {Den \number\day. \ifcase\month
\or Januar\or Februar\or M\"arz\or April\or Mai \or Juni\or
Juli\or August\or September\or Oktober \or November \or
Dezember\fi\space \number\year}}}
\def\today{{\sl {\ifcase\month
\or January\or February\or March\or April\or May \or June\or
July\or August\or September\or October \or November \or
December\fi \space\number\day,\space
                                            \number\year}}}

\newcommand{\XI}{\XI}

\newcommand{\sla}{\raise.15ex\hbox{$/$}\kern -.57em}
\newcommand{\Sla}{\raise.15ex\hbox{$/$}\kern -.70em}

\newcommand{\complex}{{\kern .1em {\raise .47ex
\hbox {$\scriptscriptstyle |$}}
    \kern -.4em {\rm C}}}
\newcommand{\real}{{{\rm I} \kern -.19em {\rm R}}}
\newcommand{\rational}{{\kern .1em {\raise .47ex
\hbox{$\scripscriptstyle |$}}
    \kern -.35em {\rm Q}}}
\renewcommand{\natural}{{\vrule height 1.6ex width
.05em depth 0ex \kern -.35em {\rm N}}}

\newcommand{\twiddle}{\lower.9ex\rlap{$\kern -.1em\scriptstyle\sim$}}

\newcommand{\eq}{\begin{equation}}
\newcommand{\eqn}[1]{\label{#1}\end{equation}}
\newcommand{\eea}{\end{eqnarray}}
\newcommand{\eqa}{\begin{eqnarray}}
\newcommand{\eqan}[1]{\label{#1}\end{eqnarray}}
\newcommand{\ba}{\begin{array}}
\newcommand{\ea}{\end{array}}
\newcommand{\eqac}{\begin{equation}\begin{array}{rcl}}
\newcommand{\eqacn}[1]{\end{array}\label{#1}\end{equation}}

\begin{document} 

\vspace{3mm}

\begin{center}

{\LARGE\bf Topological Gravity on $(D,N)-$Shift Superspace
Formulation} \vspace{3mm}
\end{center}

\begin{center}{\large
J. A. Louren\c co$^{a,b,d}$, J. A. Helay\"el Neto$^{b}$, W. Spalenza$^{b,c,d}$ } \vspace{1mm}
\noindent

$^{a}$ Universidade Federal do Esp\'{i}rito Santo, Campus S\~ao Mateus (UFES) Brazil.\\ 
$^{b}$ Centro Brasileiro de Pesquisas F\'\i sicas (CBPF) Brazil.\\
$^{c}$ Instituto Federal do Esp\'{i}rito Santo, Campus Cariacica (IFES) Brazil.\\
$^{d}$ Grupo de Pesquisa em F\'{i}sica Te\'orica (GPFT-IFES).

\end{center}

\begin{center}
\vspace{1mm} E-mails: jose.lourenco@ufes.br, helayel@cbpf.br, wesley.spalenza@ifes.edu.br
\vspace{1mm}
\end{center}

\begin{abstract}
In this contribution, we re-assess the subject of topological gravity by following the Shift
Supersymmetry formalism. The gauge-fixing of the theory goes under the
Batallin-Vilkovisky (BV) prescription based on a diagram that contains both
ghost and anti-ghost superfields, associated to the super-vielbein
and the super-Lorentz connection. We extend the formulation of the
topological gravity action to an arbitrary number of dimensions of the shift 
superspace by adopting a formulation based on the gauge-fixing for BF-type models. 
\end{abstract}

\section{Introduction}

Topological field theories have been introduced by Witten \cite{W2} and soon after applied in several areas that describe quantum-mechanical and quantum field-theoretical systems. Over the recent years, they have been applied to study non-perturbative quantum gravity \cite{Bonelli}, the issues of topological phases, spin foam, Loop Quantum Gravity, a topological approach to the cosmological constant and a number of other relevant applications. One of the basic topics for the construction of Topological Gravity are the topological Yang-Mills theories, by now fairly-well discussed and understood. However, on the other hand, the complexity of the symmetry groups and Lagrangians in topological gravity renders difficult the comparison between the results obtained by using different formalisms.
A topological gravity theory may be formulated by gauge-fixing an action that is
a topological invariant, which can be achieved by twisting an extended supergravity theory. The latter combines diffeomorphisms and local supersymmetry transformations and exhibit a considerably more complex structure whenever compared with Super-Yang-Mills theories. A good review paper on Topological Gravity, its corresponding physical observables and a number of interesting applications may be found in the work of Ref. \cite{CGLP}.

~\\
Ever since its formulation, Chern-Simons theory, treated in a supersymmetric
formalism \cite{NG,W, MZ}, has raised particular in\-te\-rest in the framework of gauge theories as a topological geometric model. Its basic geometrical objects in the principal bundle are the dreibein and the Lorentz connection, where the dreibeins form a basis for the tangent space, while the space-time metric appears as the square dreibein matrix.

~\\
The topological supersymmetrization of Yang-Mills \cite{LM} and $BF$-theories by adopting the shift formalism \cite{H,COS,BBRT,BT} provide a viable way to their extension to any superspace dimensions, with the Wess-Zumino super-gauge choice \cite{BCGLP}
used in connection with the Batallin-Vilkovisky prescription. In other words, an alternative construction can be used for the BV
diagrams, where the latter no longer accommodate fields, but, instead, superfields; they are referred to in the literature as BV Super-diagrams.

~\\
Thus, it is possible to formulate an immediate extension to an arbitrary superspace dimension
\cite{COS}. We think it is possible to proceed seemingly for topological gravity \cite{NPS}, that is, to describe this theory in a topological formalism, based on supersymmetry (SUSY, from now on): a supersymmetric
topological gravity \cite{T}. We accomplished this
construction by exploiting the shift supersymmetry formalism,
and defining the geometric supersymmetric elements of the theory from
an extension of the usual elements of the differential Riemannian formulation.
For the supersymmetrization, we define these elements in a basis super-manifold $%
{\cal M}$ with a mapping to the Euclidian-flat-space carried out by the $D-bein$. This $D-bein$ and the Lorentz connection describe the geometric sector along with the gauge-fixing fields of the theory in an Euclidean flat-space-time and the Grassmann-valued coordinates;
the latter are adjoined to the space-time coordinates to constitute the superspace of the theory.

~\\
Our starting point consists in rewriting the Chern-Simons action \cite{EWM} in the
$N=1$-shift supersymmetry formalism of a BF-type Model and, next, to carry out its complete
gauge-fixing. In the sequel, we shall propose the study of more than a single (simple) SUSY
and describe the topological gravity in four
space-time dimensions \cite{CS}. Consequently, we set up an action for
arbitrary superspace dimensions, i.e., with arbitrary $D-$ space-time dimensions
and extended $N$-SUSY.
~\\
We shall give a brief presentation of the elements of the
space-time theory. In the $D-$dimensional basis manifold $M$, we define
a metric $g_{\mu \nu }$, with index $\mu ,\,\nu, ... = 1,2,...,D $. The $D-bein$, $%
e_{\,\,\,\mu }^{a}$ allows the transformation between the basis
manifold and the Euclidean flat-space-time whose the flat-metric is
$\delta _{ab}$, with the same dimension and index $a ,\,b, ... = 1,2,...,D $. The metric is defined in both spaces,
\begin{equation}
g=g_{\mu \nu }dx^{\mu }\otimes dx^{\nu }=\delta _{ab}e^{a}\otimes
e^{b},
\end{equation}
such that $e^{a}=e_{\,\,\,\mu }^{a}dx^{\mu }$, represents the basis in the  dual $T^{*}_{p}M$, of the  tangent space $T_{p}M$, with basis elements given by $\partial_{\mu}$. 
~\\
The transformation properties between the basis manifold $M$ and flat space are
\begin{equation}
v^{\mu }=e_{\,\,\,a}^{\mu }u^{a},\,\,\,\,u^{a}=e_{\,\,\,\mu}^{a}v^{\mu}
\end{equation}
and
\begin{equation}
v_{\mu}=e_{\mu }^{\,\,\,a}u_{a},\,\,\,\,u_{a}=e_{a}^{\,\,\,\mu }v_{\mu}.
\end{equation}
where the entry $[e_{\,\,\,\mu }^{a}]$ has its inverse,
$[e_{\,\,\,a}^{\mu }]^{-1}.\,$
~\\
The connection written on the basis manifold, $\Gamma _{\,\,\,\nu \kappa
}^{\mu },$ has its correspondent in the Euclidean flat-space-time,
$\omega _{\mu \,\,b}^{\,\,\,a}=\Gamma _{\,\,\,\,\mu \lambda }^{\nu
}e_{\,\,\,\nu }^{a}e_{\,\,\,b}^{\lambda }-e_{\,\,\,b}^{\lambda
}\partial _{\mu }e_{\,\,\,\lambda }^{a},$, which is a 1-form in the basis
manifold, called Lorentz connection. Inversely, we have,
$\Gamma _{\,\,\,\,\mu \lambda }^{\nu }=e_{\,\,\,a}^{\nu }\partial
_{\mu }e_{\,\,\,\lambda }^{a}+e_{\,\,\,a}^{\nu }e_{\,\,\,\lambda
}^{b}\omega _{\mu \,\,\,b}^{\,\,\,a}.$ 

~\\
The objects that compose the dynamical sector are the curvature, $R=d\omega +\omega \wedge \omega$, and 
the torsion, $T=de+\omega \wedge e.$ These entities compose the geometrical sector
and obeys the Bianch identity relation 
\begin{equation}
D_{\omega }R=dR+\omega \wedge R =0, \,\,\,\,\,\,\,\, D_{\omega }T=dT+\omega \wedge T =0,
\end{equation}
where the covariant derivative with respect to the connection $\omega $ is $D_{\omega}\left( \cdot \right) =d\left( \cdot \right) +\omega \wedge \left(\cdot \right)$\footnote{Other notation for form covariant derivative is $D_{\omega}\left( \cdot \right) =d\left( \cdot \right) +[\omega, \left(\cdot \right)]$}. Following in the incoming Sections, we present the extension of these objects of the Maurer-Cartan-Einstein formalism to the superspace approach.

~\\
The paper is outlined as follows: in Section 2, we present a general formulation of gravity as a
super-BF model. This Section is split into five subsections to render the presentation clearer.
Section 3 presents the so-called super-BF model and an associated on-shell solution. An explicit
construction in the $D=3, N=1$-case is worked out to render manifest the whole idea of the method
we have followed. Finally, we cast our Concluding Comments in Section 4.

\section{Generalized Gravity as a super$-BF$ Model}

\subsection{Preliminary Definitions}

The supermanifold we shall work with consists of the basis manifold with the addition of
the Grassmann-valued coordinates. The supercoordinates are parametrized as 
$ z^{M}=\left( x^{\mu },\theta ^{I}\right) $ defined in a
superchart of the basis supermanifold \footnote{We stress that the space-time dimension is $D$, while $N$ is an internal label associated to the number of supersymmetries of the model;$N=1$ corresponds to a simple SUSY, $N>1$ stands for an N-extended SUSY.} ${\cal M},$ where $\theta ^{I}$ is the Grassmannian coordinates \cite{H,COS,BBRT,CCF,AR,MS} \footnote{The Grassmann Variables of the topological description may be found in \cite{H,COS,BBRT}, as well as the notations therein. For example for $N = 2$, Levi-Civita pseudo-tensor $\epsilon^{IJ}$ is the antisymmetric metric, $\epsilon^{12}=+1=-\epsilon_{12}$, where $\epsilon^{IJ}\epsilon_{JK}=\delta{^I \, _K}$ and a field or variable transforms as, $\varphi^I = \epsilon^{IJ}\varphi_{J}$ and $\varphi_I = \epsilon_{IJ}\varphi^{J}$. A scalar product is invariant under $SU(2)$ such that, $(\theta)^{2}=\frac{1}{2}\theta^I\theta_I = \frac{1}{2}\epsilon_{IJ}\theta^I\theta^J$. The derivative is defined as $\partial_I = \partial / \partial\theta^I$ and $\partial^I = \partial / \partial\theta_I$, and applied to a Grassmann coordinate gives $\partial_I\theta_J = -\epsilon_{IJ}$. Finally, the Berezin Integral is defined by $\int \theta^I=\partial_I$.}. In the Euclidean flat space-time, the coordinates are represented by $z^{A}=(x^{a},\theta ^{I})$.

~\\
The superderivatives as superforms are given by $\hat{d}=dx^{\mu }\partial _{\mu}+d\theta
^{I}\partial _{I}$. An arbitrary superfield $F(x,\theta ^{I})$, is defined by the action of the transformation
generated by the derivatives with respect to the Grassmann coordinates \cite{H,COS,BCGLP}, known as the shift operator,
and given according to what follows:
\begin{equation}
Q_{I}F(x,\theta )=\partial _{I}F(x,\theta
),\,\,\,\,\,\,\,\,\,\,\,\,I=1,...,N,
\end{equation}
where we assign fermionic supersymmetry numbers as follows: $[\theta^I]=-1$ and $[Q_I]=+1$.
~\\
In analogy to the Riemannian geometry, we set up a
superspace formalism \cite{CGLP} to treat the fundamental elements
of the description. In the SUSY formalism, we define the
$D-bein$ 1-(super)form as
\begin{equation}
\hat{E}^{a}=E_{\mu }^{a}(x,\theta )dx^{\mu }+E_{I}^{a}(x,\theta
)d\theta ^{I},  \label{1}
\end{equation}
and the 1-(super)form as
\begin{equation}
\hat{\Omega}^{ab}=\Omega _{\mu }^{ab}(x,\theta )dx^{\mu }+\Omega
_{I}^{ab}(x,\theta )d\theta ^{I}.  \label{2}
\end{equation}
In both cases, their components are form-superfields (the same being true
in all our definitions that involve superforms: superform components are
superffields).
~\\
Then, the dynamical objets of the formalism, the curvature 2-superform
and torsion 1-superform, are defined by means of the Cartan
algebra associated to the shift SUSY. The (super)curvature as 2-superform reads as
\begin{equation}
\hat{R}^{ab}=\hat{d}\,\hat{\Omega}^{ab}+\hat{\Omega}_{\,\,\,\,c}^{a}\hat{%
\Omega}^{cb}={\bf R}_{\,\,\,\,\mu \nu }^{ab}dx^{\nu }dx^{\mu }+{\bf R}%
_{I\,\,\mu }^{ab}dx^{\mu }d\theta ^{I}+{\bf R}_{IJ}^{ab}d\theta
^{I}d\theta ^{J},  \label{super-cur}
\end{equation}
where ${\bf R}^{ab}$ accommodates the genuine 2-form Riemann tensor that
we simply write as:
\begin{equation}
R^{ab}=R_{\,\,\,\,\,\,\mu \nu }^{ab}dx^{\nu }dx^{\mu }.
\label{Reimann C}
\end{equation}
Also, the supercovariant derivative is given by
\begin{eqnarray}
\hat{D}_{\hat{\Omega}}(\cdot ) &=&\hat{d}(\cdot )+[\hat{\Omega},(\cdot )] \\
&=&D_{\Omega }(\cdot )+d\theta ^{I}D_{I}(\cdot ) \\
&=&dx^{\mu }\left( \partial _{\mu }(\cdot )+[\Omega _{\mu },(\cdot
)]\right) +d\theta ^{I}\left( \partial _{I}(\cdot )+[\Omega
_{I},(\cdot )]\right)
\end{eqnarray}
and $\left( D_{\Omega }\right) _{\mu }=D_{\mu },$ is the covariant
1-form superfield derivative of the $\Omega $ connection. This yields the
(super)components of the curvature superfield:
\begin{equation}
{\bf R}^{ab}=\left( D_{\Omega }\Omega \right) ^{ab},\,\,\,\,\,\,\,\,\,\,\,%
{\bf R}_{I}^{ab}=d\Omega _{I}^{ab}+\partial _{I}\Omega
^{ab}+[\Omega ^{ac},\Omega _{I}^{cb}],\,\,\,\,\,\,\,\,\,\,\,{\bf
R}_{IJ}^{ab}=\left( D_{I}\Omega _{I}\right) ^{ab},
\label{super-curv}
\end{equation}
we can also write the curvature (\ref{Reimann C}) as $R=D_{\omega
}\omega .$

\subsection{Re-assessing $D = 3, N = 1$ and $D = 4, N = 1$ Topological Gravities}

~\\
After the papers \cite{NG,W} have appeared, a whole line of works has approached
supergravity and its topological version with the aim to accomplish
the description based on a maximally extended SUSY.
The challenge is to conveniently formulate it as a gauges theory, in view of
the huge number of degrees of freedom accommodated in all superfields. In
the formalism of shift SUSY, or topological supersymmetric
formalism, it is understood that (simple) $N=1$ superspace is considered \cite{E,H}. We
wish, in the present contribution, to formulate, with the help of the shift SUSY
\cite{H,COS,BT,BCGLP,CGLP}, gravity for any any dimensionality, $D$, and for a general
number of SUSY generators, $N$. We shall adopt the Batallin-Vilkovisky (BV) prescriptiom \cite{BV} 
combined with the Blau-Thompson minimal action gauge-fixing \cite
{BBRT,BT2}, that fix the the Lagrange multipliers associted to the gauge conditions\footnote{%
We propose the following notation for the superfield charges: $^{s\,}\Omega _{p}^{g},\,$%
where $s$ stands for the SUSY number, $g$ denotes the ghost number and $p$ indicates the form degree.}.
Finally, we need to add the Fadeev-Popov gauge-fixing action along with the BF-model term, forming the full invariant action.

~\\
Our starting point is the gravity formulation for $N=1$-SUSY
\cite{NG,CGLP,GGRS,PN,YT,DF}; for that, we define the 1-form-superfields by means of the following expansion $\hat{E}%
^{a}=E_{\mu }^{a}(x,\theta )dx^{\mu }+E_{\theta }^{a}(x,\theta )d\theta ,\,$%
as having an odd statistics, such that, the component superfields readily follow:
\begin{equation}
E^{a}=e^{a}+\theta \psi ^{a}, \label{Ea}
\end{equation}
and
\begin{equation}
E_{\theta }^{a}=\chi^{a}+\theta \phi ^{a}.  \label{super-bein-comp}
\end{equation}
The super-form $\hat{E}^{a}$ has odd statistics, because $e^{a}$
should be odd $-$ this is a 1-form with SUSY number 1 $-$ with its component $%
e_{\,\,\,\,\mu }^{a}$ being a bosonic field. To build up the model in $%
D=3 $ space-time dimensions, we define the super-connection as
\begin{equation}
\hat{\Omega}^{ab}=\Omega _{\,\,\,\,\mu }^{ab}(x,\theta )dx^{\mu
}+\Omega _{\theta \,}^{ab}(x,\theta )d\theta ,
\label{super-conec}
\end{equation}
where the components 1-form-superfields are
\begin{equation}
\Omega ^{ab}=\omega ^{ab}+\theta \varpi ^{ab},\,\,\,\,\,\Omega
_{\theta }^{ab}=\lambda ^{ab}+\theta \lambda _{\theta }^{ab}.
\label{super-conec-comp}
\end{equation}
The (super)components of the supercurvature can be expanded from expression (\ref {super-curv}).

~\\
The invariant (topological) gravity action in the dimension we are now considering is the Chern-Simons model \cite{H,BBRT,W2} in the corresponding shift superspace; therefore, here, to ensure the right field content, we need to change the $E^{a}$ given above (which would be compatible with a BF-type action) and re-write it as $E^{a}=\psi ^{a}+\theta e^{a},$ such that the integrand contains basically the expression for the gravitational Chern-Simons action:
\begin{eqnarray}
\int Q\left[ \varepsilon _{abc}E^{a}{\bf R}^{bc}\right] &=&\int
\varepsilon _{abc}[e^{a}R^{bc}-\psi ^{a}\left( D_{\omega }\varpi
\right) ^{bc}]
\nonumber \\
&=&\int d^{3}x\,\varepsilon _{abc}\varepsilon ^{\mu \nu \kappa
}\left\{ e_{\,\,\,\mu }^{a}R_{\,\,\,\,\,\,\nu \kappa }^{bc}-\psi
_{\,\,\,\mu }^{a}\left( D_{\nu }\varpi _{\kappa }\right)
^{bc}\right\} , \label{comp-n=1}
\end{eqnarray}
where $R^{ab}=d\omega ^{ab}+\omega ^{ac}\omega ^{cb}=\left(
D_{\omega }\omega \right) ^{ab}$,\thinspace $\,\,\left( D_{\omega
}\varpi \right) ^{ab}=d\varpi ^{ab}+[\omega ^{ac},\varpi ^{cb}]$
and the super-integral in this formalism is $\int d\theta =Q.$ The
final result of this supersymmetrization is an action free from the
supersymmetric charges. Then, this formulation for $D = 3$ dimensions is not interesting to our construction, because, as we have seen above, we have made a new definition of $E^{a}$, compared with (\ref{Ea}), so that the action acquires the Chern-Simons form. For that reason, we choose the BF-type formulation to give us the same expected results, and we do not need to change the form of the basic elements of the theory, $E^{a}$ for example, to adapt the supersymmetrization process.

~\\
Now, that the construction of the action for the geometric sector is
totally fixed, we need to describe the BV super-diagram \cite{COS} that
contains the super-components of the eq. (\ref{super-cur}), using the Blau-Thompson minimal action procedure
\cite{BBRT,BT}. The super-diagram in $D=3$ dimensions is given as below:
\begin{equation}
\begin{array}{c}
^{0}{\bf R}_{2}^{ab} \\
^{-1}\bar{H}_{1}^{ab}\,\,\,\,\,\,\,\,\,\,^{1}{\bf R}_{1}^{ab} \\
\,\,\,\,\,\,^{0}\bar{Z}_{0}^{ab}\,\,\,\,\,\,\,\,\,\,^{-2}\bar{W}%
_{0}^{ab}\,\,\,\,\,\,\,\,\,\,^{2}{\bf R}_{0}^{ab}
\end{array}
.  \label{diag n=1}
\end{equation}
Let us recall that this gauge-fixing procedure is a shift
gauge-fixing rather than the BRST gauge-fixing (Faddev-Popov), though it
has the similar effect on the shift degrees of freedom (shift ghosts elimination \cite{COS,CGLP}. In the some cases, this is named BRST gauge-fixing \cite{H,W2}. However, all curvature superfields and Lagrangian multipliers exhibit the same covariant BRST\ transformation: $s(\cdot)=-[C^{ab},(\cdot )],$ where $C^{ab}$ is a zero-form superfield
ghost associated to $\Omega ^{ab}.$ According to this diagram, the
action can be written as
\begin{equation}
S_{D=3}^{N=1}=\int d\theta \{^{-1}\bar{H}_{1}^{ab}\,\left( ^{0}{\bf R}%
_{2}^{ab}\right) +\,^{-2}\bar{W}_{0}^{ab}\,\left( D_{\Omega }*\,^{1}{\bf R}%
_{1}^{ab}\right) +\,^{0}\bar{Z}_{0}^{ab}\,\left( D_{\Omega }*\,^{-1}\bar{H}%
_{1}^{ab}\right) \},  \label{d=3 n=1}
\end{equation}
~\\
For $D=4$ dimensions, we can reproduce the results of (\ref{d=3 n=1}), by generalizing
the systematization \cite{E}, where the BV super-diagram is now
given by
\begin{equation}
\begin{array}{c}
^{0}{\bf R}_{2}^{ab} \\
^{-1}\bar{H}_{2}^{ab}\,\,\,\,\,\,\,\,\,\,^{1}{\bf R}_{1}^{ab} \\
\,\,\,\,\,\,^{0}\bar{Z}_{1}^{ab}\,\,\,\,\,\,\,\,\,\,^{-2}\bar{W}%
_{0}^{ab}\,\,\,\,\,\,\,\,\,\,^{2}{\bf R}_{0}^{ab} \\
^{-1}\bar{U}_{0}^{ab}\,\,\,\,\,\,\,\,\,\,\,\,\,\,\,\,\,\,\,\,\,\,\,\,\,\,\,%
\,\,\,\,\,\,\,\,\,\,\,\,\,\,\,\,\,\,\,\,\,\,\,\,\,\,\,\,\,\,\,\,\,\,\,\,\,\,%
\,\,\,\,\,\,\,\,\,
\end{array}
\label{diag d=4 n=1}
\end{equation}
In this case, we clearly notice, by means of the Blau-Thompson procedure, that the
ghost fields for each Lagrange multiplier are elimined by a
simple redefinition in the action, leading to a minimal
gauge-fixing action \cite{BBRT,BT}. Therefore, the associated action
to (\ref{diag d=4 n=1}) can be written as
\begin{eqnarray*}
S_{D=4}^{N=1} &=&\int d\theta \{^{-1}\bar{H}_{2}^{ab}\,\left( ^{0}{\bf R}%
_{2}^{ab}\right) +\,^{-2}\bar{W}_{0}^{ab}\,\left( D_{\Omega }*\,^{1}{\bf R}%
_{1}^{ab}\right) +\,^{0}\bar{Z}_{0}^{ab}\,\left( D_{\Omega }*\,^{-1}\bar{H}%
_{2}^{ab}\right) \\
&&+\,^{-1}\bar{U}_{0}^{ab}\left( D_{\Omega
}*\,^{0}\bar{Z}_{1}\right) ^{ab}\}.
\end{eqnarray*}
~\\
Now, we are ready to build up the supergravity action in $N=2$ superspace.
This task is the subject of our next Section.

\subsection{The Topological Gravity in $N=2, D=4$}
~\\
Here, it is possible to write down the topological gravity action,
using the same construction as the one in the previous Section. Now,
the $D-bein$ 1-form-superfield is given by
\begin{equation}
E^{a}=e^{a}+\theta ^{I}\psi _{I}^{a}+\frac{1}{2}\theta ^{2}\rho
^{a},\,\,\,\,\,\,\,\,E_{I}^{a}=\chi _{I}^{a}+\theta ^{J}\phi _{JI}+\frac{1}{2%
}\theta ^{2}\varphi _{I}^{a}.  \label{n=2 bein}
\end{equation}
The connection components if the 1-superform (\ref{2}) can be read as below:
\begin{equation}
\Omega ^{ab}=\omega ^{ab}+\theta ^{I}\varpi
_{I}^{ab}+\frac{1}{2}\theta ^{2}\tau ^{ab},\,\,\,\,\,\,\,\Omega
_{I}^{ab}=\lambda _{I}^{ab}+\theta ^{J}\lambda
_{JI}^{ab}+\frac{1}{2}\theta ^{2}\kappa _{I}^{ab}. \label{n=2
conec}
\end{equation}
~\\
The action for $D=3$ space-time dimensions obey the same
gauge-fixing systematization as ion the $N=1$ case. This super-diagram here
can be written as
\begin{equation}
\begin{array}{c}
^{0}{\bf R}_{2}^{ab} \\
^{-2}\bar{H}_{1}^{ab}\,\,\,\,\,\,\,^{1}{\bf R}_{1\,\,I}^{ab} \\
\,\,\,\,\,\,^{0}\bar{Z}_{0}^{ab}\,\,\,\,\,\,\,^{-3}\bar{W}%
_{0}^{ab\,\,I}\,\,\,\,\,\,\,^{2}{\bf R}_{0\,\,\,IJ}^{ab}
\end{array}
,
\end{equation}
For the invariant topological gravity model in $D=4$ dimensions we propose

\begin{equation}
\begin{array}{c}
^{0}{\bf R}_{2}^{ab} \\
^{-2}\bar{H}_{2}^{ab}\,\,\,\,\,\,\,\,^{1}{\bf R}_{1\,\,\,I}^{ab} \\
\,\,\,\,\,\,^{0}\bar{Z}_{1}^{ab}\,\,\,\,\,\,^{-3}\bar{W}_{0}^{ab\,\,I}\,\,\,%
\,\,\,\,^{2}{\bf R}_{0\,\,\,IJ}^{ab} \\
^{-2}\bar{U}_{0}^{ab}\,\,\,\,\,\,\,\,\,\,\,\,\,\,\,\,\,\,\,\,\,\,\,\,\,\,\,%
\,\,\,\,\,\,\,\,\,\,\,\,\,\,\,\,\,\,\,\,\,\,\,\,\,\,\,\,\,\,\,\,\,\,\,\,\,\,%
\,\,\,\,\,\,\,\,\,
\end{array}
,
\end{equation}
Therefore the action associated to this diagram is given as
\begin{eqnarray}
S_{D=4}^{N=2} &=&\int d^{2}\theta \{\,^{-2}\bar{H}_{2}^{ab}\,\left( ^{0}{\bf %
R}_{2}^{ab}\right) +\,^{-3}\bar{W}_{0}^{ab\,\,I}\,\left( D_{\Omega }*\,^{1}%
{\bf R}_{1\,\,\,\,I}^{ab}\right)  \nonumber \\
&&+\,^{0}\bar{Z}_{1}^{ab}\,\left( D_{\Omega
}*\,^{-2}\bar{H}_{2}^{ab}\right) +\,^{-2}\bar{U}_{0}\,\left(
D_{\Omega }*\,^{0}\bar{Z}_{1}^{ab}\right) \},
\end{eqnarray}
~\\
Once we have understood how an action can be written, we shall define the
geometric elements for a general superspace dimension and we shall
present the action of topological gravity in the sequel.

\subsection{The $(N,D)-$Superspace}
~\\
We generalize the formulation by expanding the action from $N=2$ to an
arbitrary number of SUSYs. We need to define the form-superfields for $N$
superspace dimensions. The shift index runs as $I=1,...,N,$ and the
Levi-Civita pseudo-tensor is defined
as: $\epsilon ^{1...N}=1$. The $D-bein$ (super)component fieds of eq. (%
\ref{1}) are 1-form superfields and their expansion in component
fields is given by
\begin{eqnarray}
E^{a} &=&e^{a}+\theta ^{I}\psi _{I}^{a}+...+\frac{1}{N!}\theta
^{N}\psi
_{N}^{a}, \\
E_{I}^{a} &=&\phi _{I}^{a}+\theta ^{J}\phi
_{JI}^{a}+...+\frac{1}{N!}\theta ^{N}\varphi _{I}^{a}.
\end{eqnarray}
The component connections of the 2-superform (\ref{2}) are
\begin{eqnarray}
\Omega ^{ab} &=&\omega ^{ab}+\theta ^{I}\omega _{I}^{ab}+...+\frac{1}{N!}%
\theta ^{N}\varpi ^{ab}, \\
\Omega _{I}^{ab} &=&\lambda _{I}^{ab}+\theta ^{J}\lambda _{IJ}^{ab}+...+%
\frac{1}{N!}\theta ^{N}\varrho _{I}^{ab}.
\end{eqnarray}
~\\
In $D=3$ dimensions, the construction is the same for all $N$,
i.e., as (\ref {diag n=1})$.$ In $D=4$ dimensions, the diagram is
also similar to (\ref{diag d=4 n=1}). We can extend the space-time
dimension  from $4$ to any $D$. We start off the
construction with the Levi-Civita symbol, where we assume $\varepsilon
^{012...D-1}=1.$ The the Euclidian flat-space-time indices are
$a_{1},a_{2},...,a_{D}$ and they run from $1$ to $D$. The BV
super-diagram in the $(D,N)-$superspace is
\begin{equation}
\begin{array}{c}
^{0}{\bf R}_{2}^{ab} \\
^{-N}\bar{H}_{D-2}^{ab}\,\,\,\,\,\,^{1}{\bf R}_{1\,\,\,I}^{ab} \\
^{0}\bar{Z}_{D-3}^{ab}\,\,\,^{-N-1}\bar{W}_{0}^{ab\,\,I}\,\,\,\,^{2}{\bf R}%
_{0\,\,\,IJ}^{ab} \\
^{-N}\bar{Z}_{D-4}^{ab}\,\,\,\,\,\,\,\,\,\,\,\,\,\,\,\,\,\,\,\,\,\,\,\,\,\,%
\,\,\,\,\,\,\,\,\,\,\,\,\,\,\,\,\,\,\,\,\,\,\,\,\,\,\,\,\,\,\,\,\,\,\,\,\,\,%
\,\,\,\,\,\,\,\,\,\,\,\,\,\, \\
\swarrow
\,\,\,\,\,\,\,\,\,\,\,\,\,\,\,\,\,\,\,\,\,\,\,\,\,\,\,\,\,\,\,\,\,\,\,\,\,\,%
\,\,\,\,\,\,\,\,\,\,\,\,\,\,\,\,\,\,\,\,\,\,\,\,\,\,\,\,\,\,\,\,\,\,\,\,\,\,%
\,\,\,\,\,\,\,\,\,\,\,\,\,\,\,\,\,\,\, \\
^{-s}\bar{Z}_{0}^{ab}\,\,\,\,\,\,\,\,\,\,\,\,\,\,\,\,\,\,\,\,\,\,\,\,\,\,\,%
\,\,\,\,\,\,\,\,\,\,\,\,\,\,\,\,\,\,\,\,\,\,\,\,\,\,\,\,\,\,\,\,\,\,\,\,\,\,%
\,\,\,\,\,\,\,\,\,\,\,\,\,\,\,\,\,\,\,\,\,\,\,\,\,\,\,\,\,\,\,\,\,\,\,\,\,\,%
\,\,\,\,\,\,\,\,\,\,\,\,\,\,\,
\end{array}
\label{diag DN}
\end{equation}
with $s=0$ if $D$ even and $s=-N\,$ if $D$ odd. Therefore, the
invariant action in the $(D,N)-$superspace is
\begin{eqnarray}
S_{D}^{N} &=&\int d^{N}\theta \{^{-N}\bar{H}_{D-2}^{ab}\,\,^{0}{\bf R}%
_{2}^{ab}+\,^{-N-1}\bar{W}_{0}^{ab\,\,I}\,\left( D_{\Omega }*\,^{1}{\bf R}%
_{1}^{ab}\right)   \nonumber \\
&&+\,^{0}\bar{Z}_{D-3}^{ab}\,\left( D_{\Omega }*\,^{-N}\bar{H}%
_{D-2}^{ab}\right) +\,^{-N}\bar{Z}_{D-4}^{ab}\,\left( D_{\Omega }*\,^{0}\bar{%
Z}_{D-3}^{ab}\right)   \nonumber  \label{tot act} \\
&&+...+\,^{-s}\bar{Z}_{0}^{ab}\,\left( D_{\Omega }*\,^{-N-s}\bar{Z}%
_{1}^{ab}\right) \},
\end{eqnarray}
where each $N-$power covariant derivative is written as: $D_{\Omega
_{I}}^{N}=\epsilon ^{I_{1}...I_{N}}D_{I_{1}}...D_{I_{N}}.$
~\\
This action still presents still spurious degrees of freedom; they are localized
in the connection $\Omega _{\mu }(x,\theta )$; thus, we need to
suppress them from the model by fixing the gauge with the help of the BRST ghosts,
as we are next going to do.

\subsection{The Fadeev-Popov Gauge-Fixing}
~\\
According to \cite{H,COS}, the gauge-fixing of topological
gravity consists in writing the Batallin-Vilkovisky diagram
\cite{CGLP} and defining the respective ghost for each field. The ghost field 
and the diffeomorphism transformation fix the principal action \cite{PS}. Here, we need to fix
only the Lorentz gauge condition. The ghost superfield, for $N=1$-SUSY, for example, is
\begin{equation}
\hat{C}^{ab}=C^{ab}=c^{ab}+\theta c_{\theta }^{ab},
\end{equation}
Associated to this ghost, we need to define the anti-ghosts and their
Lagrange multipliers as zero-form-superfields,
\begin{equation}
\bar{C}^{ab}=\bar{c}^{ab}+\theta \bar{c}_{\theta
}^{ab},\,\,\,\,\,\,\,\,\,\,\,\,B^{ab}=b^{ab}+\theta b_{\theta
}^{ab}.
\end{equation}
The diffeomorphism ghost field is a super-vector in shift
superspace \cite{CGLP}, such that,
\[
\hat{\Xi}=\Xi ^{M}\partial _{M}=\Xi ^{\mu }(x,\theta )\partial
_{\mu }+\Xi ^{I}(x,\theta )\partial _{I};
\]
its component fields read as follows:
\[
\Xi ^{\mu }=\xi ^{\mu }+\theta \xi _{\theta }^{\mu
},\,\,\,\,\,\,\,\Xi ^{\theta }=\zeta +\theta \zeta _{\theta },
\]
and this superform obeys the following properties
\begin{eqnarray}
\hat{\Xi}^{2} &=&\frac{1}{2}[\hat{\Xi},\hat{\Xi}],\,\,\,\,\,\,\,\,\,\,\,\,\,%
\,(\hat{\Xi}^{2})^{M}=\left( \Xi ^{N}\partial _{N}\right) \Xi ^{M}, \\
\pounds _{\hat{\Xi}} &=&[i_{\hat{\Xi}},\hat{d}],\,\,\,\,\,\,\,\,\,\,\,\,\,%
\,[\pounds _{\hat{\Xi}_{1}},\pounds _{\hat{\Xi}_{2}}]=\pounds _{[\hat{\Xi}%
_{1},\hat{\Xi}_{2}]}.
\end{eqnarray}
where $\pounds _{\hat{\Xi}}$ is the Lie derivative and $i_{\Xi}$ denotes the inner product operation \cite{CGLP}.
~\\
The BRST transformation consists of the ghost and the
diffeomorphism transformations. Those transformations on the
super-forms described in the equations, (\ref{2}), are given by
\begin{eqnarray}
s\hat{\Omega}^{ab} &=&\pounds _{\hat{\Xi}}\hat{\Omega}^{ab}-\left( \hat{D}_{%
\hat{\Omega}}C\right) ^{ab}, \\
sC^{ab} &=&\pounds _{\hat{\Xi}}C^{ab}-\left( C*C\right) ^{ab}, \\
s\bar{C}^{ab} &=&\pounds _{\Xi }\bar{C}^{ab}+B^{ab}, \\
\,sB^{ab} &=&0, \\
s\hat{\Xi} &=&\hat{\Xi}^{2}.
\end{eqnarray}
where the BRST operator is nilpotent, $s^{2}=0,$ and anticommutes with $\hat{d%
}.$ From the complete set of the BRST transformations described
above, we can write the super-diagram of the Batallin-Vilkovisky to
fix the spurious degrees of freedom of the Lorentz connection. The
super-diagram associated to the general action (\ref{tot act}),
will be
\begin{equation}
\begin{array}{c}
\Omega ^{ab} \\
\bar{C}^{ab}\,\,\,\,\,\,\,\,\,C^{ab}
\end{array}
.  \label{diag-fp-old}
\end{equation}
The Fadeev-Popov gauge-fixing super-Lagrangian, invariant under
the BRST operator, is:
\begin{equation}
{\cal L}_{gf}=s\left\{ \bar{C}^{ab}\left( D_{\Omega }*\Omega
\right) ^{ab}\right\} ,
  \end{equation}
so that, according to the BRST transformation above, we have
the action
\begin{eqnarray}
S_{gf} &=&\int d\theta \{B^{ab}\left( D_{\Omega }*\Omega \right) ^{ab}-%
\bar{C}^{ab}\left( D_{\Omega }*D_{\Omega }C\right) ^{ab}  \nonumber \\
&&+\pounds _{\Xi }\bar{C}^{ab}\left( D_{\Omega }*\Omega \right) ^{ab}+\bar{C}%
^{ab}\left( D_{\Omega }*\pounds _{\Xi }\Omega \right) ^{ab}\},
\label{gf}
\end{eqnarray}
where "$*$" is the dual star transformation, or star product, of a object from a tangent space to a cotangent space. This action in one Grassmann dimension has the same structure, in the superfield form, of the action for N-dimensions, because the star product guarantee this notation.
Therefore, the invariant gauge-fixed gravity action for any dimension is determined by the sum of
(\ref{tot act}) and (\ref{gf}); this yields
\begin{equation}
S=S_{D}^{N}+S_{gf}.  \label{tot-old}
\end{equation}

\section{The super$-BF$ Model and an On-shell Solution}
~\\
The model studied in the previous Section is
acceptable, because it correctly describes the particular models previousloy studied; 
nevertheless, we can formulate a new model whose geometric
sector in the action is stable under the BRST invariance of all the other sectors, which
we cannot guarantee in the general model of the previous Section. Here, we do not follow
any longer a Blau-Thompson minimal action procedure. The diference
consists basically in considering the total action written as a BRST
invariance, considering also the on-shell instanton solutions of
the geometric sector. This means that the first Lagrangian multiplier contains
the background zero-modes of the degrees of freedom. Therefore,
by virtue of the BRST transformation for the first Lagrangian
multiplier $^{-N}\bar{H}_{D-2}^{ab}$ of the super-diagram
(\ref{diag DN}), the latter gets restricted to the geometric section and the
first multipliers according to what follows:
\begin{equation}
\begin{array}{c}
^{0}{\bf R}_{2}^{ab} \\
^{-N}\bar{H}_{D-2}^{ab}\,\,\,\,\,\,^{1}{\bf R}_{1\,\,\,I}^{ab} \\
\,\,\,\,\,\,\,\,\,\,\,\,\,\,\,\,\,\,\,\,\,\,\,\,^{-N-1}\bar{W}%
_{0}^{ab\,\,I}\,\,\,\,\,\,\,^{2}{\bf R}_{0\,\,\,IJ}^{ab}
\end{array}
,  \label{diag-new}
\end{equation}
and the rest of the gauge-fixing diagram splits from the one above;
it involves only the fixation of the $\bar{H}^{ab}$and its
anti-ghosts. The BRST transformation of the Lagrangian multiplier
is then
\begin{equation}
s\,\bar{H}^{ab}=-D_{\Omega }\,\Sigma
^{ab}-[C^{ac},\,\bar{H}^{cb}]+\pounds _{\Xi }\,\bar{H}^{ab}.
\label{brst}
\end{equation}
For the other superfields of the super-diagram (\ref{diag-new}),
the BRST\ transformation is covariant: $s(\cdot )=-[C,(\cdot )].$
To preserve the nilpotency of the BRST operator $s,$ we take the
on-shell solution for the new model as an instanton solution,
${\bf R}^{ab}-*{\bf R}^{ab}=0$ in four dimensions$,$ and for
other dimensionalities the null curvature solution, ${\bf
R}^{ab}=0$. We now write the BV super-diagram associated to this
Lagrangian multiplier:
\begin{equation}
\begin{array}{c}
^{-N}\bar{H}_{D-2}^{0} \\
^{0}\bar{\Sigma}_{D-3}^{-1}\,\,\,^{-N}\Sigma _{D-3}^{1} \\
^{-N}\bar{\Sigma}_{D-4}^{0}\,\,\,^{0}\bar{\Sigma}_{D-4}^{-2}\,\,^{-N}\Sigma
_{D-4}^{2} \\
\swarrow \,\,\,\,\,\,\,\,\,\,\,\,\,\,\,\,\,\,\,\,\,\,\,\cdots
\,\,\,\,\,\,\,\,\,\,\,\,\,\,\,\,\,\,\,\,\,\,\,\,\searrow \\
^{s}\bar{\Sigma}_{0}^{g}\,\,\,\,\,\,\,\,\,\,\,\,\,\,\,\,\,\,\,\,\,\,\,\,\,\,%
\,\,\,\,\,\,\,\,\,\cdots
\,\,\,\,\,\,\,\,\,\,\,\,\,\,\,\,\,\,\,\,\,\,\,\,^{-N}\Sigma
_{0}^{D-2}
\end{array}
,  \label{fp}
\end{equation}
whose the transformations for all ghosts and anti-ghosts read as
\begin{equation}
s\left( ^{s}\,\Phi \,_{D-2-g}^{-(-1)dig(g)}\right) =-D_{\Omega
}\left( ^{s}\,\Phi \,_{D-3-g}^{-(-1)dig(g+1)}\right)
-[C,^{s}\,\Phi \,_{D-2-g}^{-(-1)dig(g)}]+\pounds
_{\hat{\Xi}}\left( ^{s}\,\Phi \,_{D-2-g}^{-(-1)dig(g)}\right) ,
\label{new gau}
\end{equation}
where $\Phi =\{^{s}\Sigma _{p}^{g},^{s}\bar{\Sigma}_{p}^{g}\}.$
The super-diagram of the BRST Lagrange multipliers of the anti-ghosts $^{s}\bar{%
\Sigma}_{p}^{g}$ is
\begin{equation}
\begin{array}{c}
^{0}\Pi _{D-3}^{0} \\
^{-N}\Pi _{D-4}^{1}\,^{0}\Pi _{D-4}^{1} \\
\cdots \,\,\,\,\,\,\,\,\,\,\,\,\,\cdots
\,\,\,\,\,\,\,\,\,\,\,\cdots
\end{array}
,  \label{fp1}
\end{equation}
with
\begin{equation}
s\,\,^{s}\bar{\Sigma}_{p}^{g}=\pounds _{\Xi }\,^{s}\bar{\Sigma}%
_{p}^{g}+\,^{s}\Pi
_{p}^{g-1},\,\,\,\,\,\,\,\,\,\,\,\,\,\,\,s^{s}\Pi _{p}^{g-1}=0.
\end{equation}
$\,$We define only the anti-fields $\bar{H}^{*},\,\,\,\bar{W}^{*}$
of the Lagrangian multipliers $\bar{H},\,\,\bar{W}$ in the diagram
(\ref{diag-new}) that fix the geometric sector. Those describe the
following relation
\begin{eqnarray}
s_{\ell }\int d^{N}\theta \left( \bar{H}^{*}\,\bar{H}+\bar{W}^{*}\,\bar{W}%
\right) &=&\left\{ \left( {\cal N}_{\bar{H}}-{\cal
N}_{\bar{H}^{*}}\right) +\left( -{\cal N}_{\bar{W}}+{\cal
N}_{\bar{W}^{*}}\right) \right\} S_{Geom}
\nonumber \\
&=&\int d^{N}\theta \left[ \bar{H}^{ab}{\bf
R}^{ab}+\bar{W}^{ab\,I}\left( D_{\Omega }*{\bf R}_{I}^{ab}\right)
\right]
\end{eqnarray}
where the geometric sector action is: $S_{Geom}=\int d^{N}\theta \left[ \bar{%
H}\,{\bf R}+\bar{W}^{\,I}\left( D_{\Omega }*{\bf R}_{I}\right) \right] .\,$%
The$\,$ ${\cal N}$ is the counter field operator and the
Slavnov-Taylor operator is given by
\begin{equation}
{\cal S}=s_{\ell }+{\cal O}(\hbar ),
\end{equation}
with the following properties
\begin{eqnarray}
{\cal S}_{S}{\cal S}(S)
&=&0,\,\,\,\,\,\,\,for\,\,\,\,\,\,\,\forall
\,\,\,\,\,S, \\
{\cal S}_{S}{\cal S}_{S}
&=&0,\,\,\,\,\,\,\,if\,\,\,\,\,\,\,\,{\cal S}(S)=0.
\end{eqnarray}
and $s_{\ell }$ is the linearized BRST operator.

~\\
The invariant gauge action is determined by the geometric sector
action for the super-diagram (\ref{diag-new}) and the Fadeev-Popov
action of the super-diagrams (\ref{diag-fp-old}), (\ref{fp}),
(\ref{fp1}), such that
\begin{eqnarray}
S &=&\int d^{N}\theta \{\,^{-N}\bar{H}_{D-2}^{ab}\,{\bf R}^{ab}+\,^{-N-2}%
\bar{W}_{0}^{ab\,\,I}\,\left( D_{\Omega }*{\bf R}_{I}^{ab}\right)
\nonumber
\\
&&+s\,\,[\,\,^{0}\overline{\Sigma }_{D-3}^{-1}(D_{\Omega }*\,^{-N}\bar{H}%
_{D-2}^{0})+\,^{0}\bar{\Sigma}_{D-4}^{-2}\,(D_{\Omega
}*\,^{-N}\Sigma
_{D-3}^{1})  \nonumber \\
&&+\,^{-N}\bar{\Sigma}_{D-4}^{0}\,(D_{\Omega }*\,^{0}\Sigma
_{D-3}^{-1})+...+\,^{s}\Sigma _{0}^{g}\,(D_{\Omega
}*\,^{-s-1}\Sigma
_{0}^{-g-1})  \nonumber \\
&&+\bar{C}^{ab}\left( D_{\Omega }*\Omega \right) ^{ab}]\,\},
\label{tot-new}
\end{eqnarray}
where $s=-N$ \thinspace and$\,\,g=0$ if $D$ is even or $s=0$ and $g=-1$ if $%
D\,$\thinspace is odd. Summarizing, the general form of this new
model (\ref {tot-new}) exhibits trivial cohomology for both
operators $Q$ and $s$,
\begin{equation}
S=Q\,\,s\,({\cal L}_{Geometric}+{\cal L}_{Fadeev-Popov}).
\end{equation}
while in the previous Section the total action (\ref{tot-old})
obeys the scheme
\begin{equation}
S=Q({\cal L}_{Geometric}+s\,{\cal L}_{Fadeev-Popov})
\end{equation}

~\\
For the sake of illustrating the gauge-fixing procedure, we can present a simple example in the particular $%
D=3$ and $N=1$ case.

{\bf Example:\ }Let us reproduce the topological gravity in the case of three
dimensions with $N=1$-SUSY$.$ The super-diagram is
\begin{equation}
\begin{array}{c}
^{0}{\bf R}_{2}^{ab} \\
^{-1}\bar{H}_{1}^{ab}\,\,\,\,\,\,\,\,^{1}{\bf R}_{\theta \,\,\,1}^{ab} \\
\,\,\,\,\,\,\,\,\,\,\,\,\,\,\,\,\,\,\,\,\,\,\,^{-2}\bar{W}%
_{0}^{ab}\,\,\,\,\,\,\,^{2}{\bf R}_{\theta \theta \,\,\,0}^{ab}
\end{array}
.
\end{equation}
Using the defintions (\ref{super-conec}),
(\ref{super-conec-comp}), the curvature
(\ref{super-curv}) and the Lagrange multiplier superfields are given,
component-wise, as follows:
\begin{eqnarray}
{\bf R}^{ab} &=&R^{ab}-\theta \left( D_{\omega }\varpi \right) ^{ab}, \\
{\bf R}_{\theta }^{ab} &=&\varpi ^{ab}+\left( D_{\omega }\lambda
\right) ^{ab}+\theta \left( \varpi ^{ac}\lambda ^{cb}-\left(
D_{\omega }\lambda
_{\theta }\right) ^{ab}\right) , \\
\bar{H}^{ab} &=&h^{ab}+\theta h_{\theta }^{ab}, \\
\bar{W}^{ab} &=&w^{ab}+\theta w_{\theta }^{ab}.
\end{eqnarray}
so that, all those superfields are covariant under the BRST transformation, $%
s(\cdot )=-[C,(\cdot )]$. The BRST transformation for the Lagrange
multipliers contains the zero-modes of the degrees of freedom given in the
eq. (\ref {brst})$.$ We define the BRST ghost and anti-ghost as
zero-form superfields
\begin{eqnarray}
\Sigma ^{ab} &=&\sigma ^{ab}+\theta \sigma _{\theta }^{ab}, \\
\bar{\Sigma}^{ab} &=&\bar{\sigma}^{ab}+\theta \bar{\sigma}_{\theta
}^{ab}.
\end{eqnarray}
The BRST transformation for all superfields of the Fadeev-Popov
gauge-fixing action is given by
\begin{eqnarray}
s\hat{\Omega}^{ab} &=&\pounds _{\hat{\Xi}}\hat{\Omega}^{ab}-\left( \hat{D}_{%
\hat{\Omega}}C\right) ^{ab}, \\
sC^{ab} &=&\pounds _{\hat{\Xi}}C^{ab}-\left( C*C\right) ^{ab}, \\
s\bar{C}^{ab} &=&\pounds _{\hat{\Xi}}\bar{C}^{ab}+B^{ab} \\
sB^{ab} &=&0,
\end{eqnarray}
the transformation for the shift anti-ghost sector carries a new
term: $\left( D_{\Omega }\Sigma \right) ^{ab}$ discussed in
(\ref{brst}) such that
\begin{equation}
s\,\bar{H}^{ab}=\pounds _{\hat{\Xi}}\bar{H}^{ab}-[C^{ac},\bar{H}%
^{cb}]-\left( D_{\Omega }\Sigma \right) ^{ab}  \label{new gauge
tras}
\end{equation}
and the other transformation is given by
\begin{eqnarray}
s\Sigma ^{ab} &=&\pounds _{\hat{\Xi}}\Sigma ^{ab}-\left( \Sigma
*\Sigma
\right) ^{ab}, \\
s\bar{\Sigma}^{ab} &=&\pounds _{\hat{\Xi}}\bar{\Sigma}^{ab}+\Pi ^{ab}, \\
s\Pi ^{ab} &=&0,\, \\
s\,\hat{\Xi} &=&\hat{\Xi}\,^{2}.
\end{eqnarray}
For the super-diagram of the Lagrange multiplier of ${\bf R}^{ab}$
adapted for the $N=1$ case, have
\begin{equation}
\begin{array}{c}
\bar{H}^{ab} \\
\bar{\Sigma}^{ab}\,\,\,\,\,\,\,\,\,\,\Sigma ^{ab}
\end{array}
,
\end{equation}
Thus, the Fadeev-Popov gauge-fixing super-Lagrangian, together
with (\ref {diag-fp-old}), becomes
\begin{equation}
s\left\{ \bar{C}^{ab}\left( D_{\Omega }*\Omega \right) ^{ab}+\bar{\Sigma}%
^{ab}\left( D_{\Omega }*\bar{H}\right) ^{ab}\right\} .
\end{equation} 
According to the eq. (\ref{tot-new}) and the BRST transformation
above, the explicit topological supergravity invariant action 
takes the form
\begin{eqnarray}
S &=&\int d\theta \{\bar{H}^{ab}{\bf R}^{ab}+\bar{W}^{ab}{\bf
R}_{\theta
}^{ab}  \nonumber \\
&&+B^{ab}\left( D_{\Omega }*\Omega \right)
^{ab}-\bar{C}^{ab}\left(
D_{\Omega }*D_{\Omega }C\right) ^{ab}  \nonumber \\
&&+\Pi ^{ab}\left( D_{\Omega }*\bar{H}\right)
^{ab}-\bar{\Sigma}^{ab}\left(
D_{\Omega }*D_{\Omega }\Sigma \right) ^{ab}  \nonumber \\
&&+\pounds _{\hat{\Xi}}\bar{C}^{ab}\left( D_{\Omega }*\Omega \right) ^{ab}+%
\bar{C}^{ab}\left( D_{\Omega }*\pounds _{\hat{\Xi}}\Omega \right)
^{ab}
\nonumber \\
&&+\pounds _{\hat{\Xi}}\bar{\Sigma}^{ab}\left( D_{\Omega
}*\bar{H}\right) ^{ab}+\bar{\Sigma}^{ab}\left( D_{\Omega }*\pounds
_{\hat{\Xi}}\bar{H}\right)
\\
&&-\bar{\Sigma}^{ab}\left( D_{\Omega }*[C,\bar{H}]\right) ^{ab}.
\end{eqnarray}
According to the new term of the transformation (\ref{new gauge
tras}), this requires that the on-shell solution for the
superfield curvature be null because we work in a 3-dimensional
manifold.

~\\
A good example is the topological gravity theory for a
two-dimensional Riemannian world-sheet manifold, that usually appears coupled
to topological sigma-models. The dimension constrains the
connection and curvature superforms not to carry the
Euclidean vector index; they should be represented as $\hat{\Omega}$ and $\hat{R}$.
For the BV super-diagram, we may follow the same systematic construction of
example contemplated above, but the Lagrange multiplier associated to ${\bf R}$ is a
zero-form. For that reason, it does not need to have an associated anti-ghost; thus, the
gauge-fixing systematic procedure does not change.

\section{Concluding Comments}
~\\
Based on the investigation pursued here, we are able to write down general models over Riemannian manifolds
endowed with a metric, both in the topological supersymmetric
and in the Witten-type topological formulations, preserving topological invariants like the Euler
characteristic, $\chi $, and the correlation functions, while keeping them free from shift spurious degrees of
freedom. A good prospective for the application of our results would be the association with
twist techniques to re-obtain ordinary supergravity theory in the Weyl representation. This attempt, by using
topological supergravity, can be implemented for any superspace dimension because there are no symmetry limitations
in superspace that prevents us from following this path. Other possibilities of apllications that we may point out 
could be in the framework of Loop Quantum Gravity, Spin foam, Global effects with continuous deformations and systems
with emergent SUSY, like, for instance, interesting topological materials such as Weyl semi-metals and topological
insulators. Finally, an open issue which shall be the subject of a forthcoming work is the coupling of the matter sector
in the framework of topological gravity. We shall be soon reporting on that.

~\\
The authors state hereby that there is no conflict of interest regarding the publication of this paper.



\begin{thebibliography}{99} 

\bibitem{W2}  E. Witten, Commun. Math. Phys. {\bf 117} (1988) 353; Int. J. Mod. Phys. {\bf A6} (1991) 2775. 

\bibitem{Bonelli} G. Bonelli and A.M. Boyarski, “Six dimensional topological gravity and the
cosmological constant problem”, Phys.Lett. B490 (2000) 147, hep-th/0004058; \\
A. Perez, “Spin foam models for quantum gravity”, Class.Quant.Grav. 20
(2003) R43, gr-qc/0301113; \\
T. Thiemann, “Lectures on loop quantum gravity”, in Quantum Gravity: From
Theory to Experimental Search, D.J.W. Giulini, C. Kiefer and C. L¨ammerzahl,
eds., Lecture Notes in Physics Vol.631 (Springer Verlag, 2003), gr-qc/0210094.

\bibitem{CGLP}  C. P. Constantinidis, A. Deandrea, F. Gieres, M. Lefran\c{c}oies and O. Piguet, Class. Quant. Grav. {\bf 21} (2004) 3515.

\bibitem{NG}  H. Nishino and S. J. Gates Jr., Int. J. Mod. Phys. {\bf A8}, 3371
(1993).

\bibitem{W}  E. Witten, Nucl. Phys. {\bf B311}, 46, (1988/89). 

\bibitem{MZ}  M. Hassaine and J. Zanelli, {\it Chern Simons (Super)Gravity (100 Years of General Relativity)} World Scientific Publishing (2016). 

\bibitem{LM}  J. Labastida and M. Marino, {\it Topological Quantum Field Theory and Four Manifolds}, Ma\-the\-ma\-ti\-cal Physics Studies (2005).

\bibitem{H}  J.H. Horne, Nucl. Phys.\ {\bf B318} (1989) 2234.

\bibitem{COS}  C. P. Constantinidis, O. Piguet, W. Spalenza, Eur. Phys. J. {\bf C33} 444 (2004). 

\bibitem{BBRT}  D. Birmingham, M. Blau, M. Rakowski and G. Thompson, Phys. Rep. {\bf 209} (1991) 129. 

\bibitem{BT}  M. Blau and G. Thompson, Nucl.Phys. {\bf B492} (1997) 545. 

\bibitem{BCGLP}  J.L. Boldo, C.P. Constantinidis, F. Gieres, M. Lefran\c{c}ois and O. Piguet, Int. J. Mod. Phys. {\bf A19} (2004) 2971; Int. J. Mod. Phys. {\bf A18} (2003) 2119. 

\bibitem{NPS} N. Merino, A. Perez, P. Salgado, Phys. Lett. {\bf B681} (2009) 85.

\bibitem{T}  A. Toon, Class. Quant. Grav. {\bf 14} (1997) 915. 

\bibitem{EWM} E. W. Mielke, {\it Geometrodynamics of Gauge Fields: On the Geometry of Yang-Mills and Gra\-vi\-ta\-tio\-nal Gauge Theories}, Mathematical Physics Studies  (2017). 

\bibitem{CS} P. Catalána, F. Izaurieta, P. Salgado and S. Salgado, Phys. Lett. B {\bf 751} (2015) 205.

\bibitem{CCF} C. Carmeli, L. Caston and R. Fioresi, {\it Mathematical Foundations of Supersymmetry}, European Mathematical Society (2011).

\bibitem{AR} A. Rogers, {\it Supermanifolds: Theory and Applications}, World Scientific Publishing (2007). 

 

\bibitem{E}  K. Ezawa, Prog. Theor. Phys. {\bf 95} (1986) 863.

\bibitem{BV}  I.A. Batalin and G.A. Vilkovisky, Phys. Lett. {\bf B69} (1977)
309; Phys. Lett. {\bf B102} (1981) 27; Phys. Rev. {\bf D28} (1983) 2567. 

\bibitem{BT2}  M. Blau and G. Thompson, Comm. Math. Phys. {\bf 152} (1993)
41. 

\bibitem{GGRS}  E. B. Manoukian, {\it Quantum Field Theory II: Introductions to Quantum Gravity, Supersymmetry and String Theory }, Graduate Texts in Physics (2016).

\bibitem{PN} P. Nath, {\it Supersymmetry, Supergravity, and Unification},
First Edition, Cambridge Monographs on Mathematical Physics (2017). 

\bibitem{YT} Y. Tanii, {\it Introduction to Supergravity}, Springer Briefs in Mathematical Physics (2014). 

\bibitem{DF}  D. Z. Freedman, A. V. Proeyen, {\it Supergravity}, First Edition, Cambridge University Press (2012).

\bibitem{MS}  P. de Medeiros and B. Spence,\ Class. Quant. Grav. {\bf 20}
(2003) 2075.


\bibitem{PS}  O. Piguet and S.P. Sorella, {\it Algebraic Renormalization},
Springer-Verlag (1995).



\end{thebibliography}
\end{document}